\documentclass[conference]{IEEEtran}

\usepackage[linesnumbered,ruled,vlined]{algorithm2e}
\usepackage{algorithm2e}

\usepackage{cite}
\usepackage{hyperref}
\usepackage{amsmath,amssymb,amsfonts}
\usepackage{algorithmic}
\usepackage{graphicx}
\usepackage{textcomp}
\usepackage{xcolor}
\usepackage{multirow}
\usepackage{booktabs}
\usepackage[caption=false]{subfig}

\def\BibTeX{{\rm B\kern-.05em{\sc i\kern-.025em b}\kern-.08em
    T\kern-.1667em\lower.7ex\hbox{E}\kern-.125emX}}
\usepackage[labelformat=parens,labelsep=quad,skip=3pt]{caption}
\usepackage{graphicx}
\begin{document}
\newcommand{\smalltrain}{{\sc TrainInGPU}\xspace}
\newcommand{\coa}{{\sc MultiEdgeCollapse}\xspace}
\newcommand{\malgo}{{\sc Gosh}\xspace}
\newcommand{\graphvite}{{\sc \emph{GraphVite}}\xspace}
\newcommand{\grappolo}{{\sc \emph{Grappolo}}\xspace}
\newcommand{\Line}{{\sc \emph{LINE}}\xspace}
\newcommand{\nodevec}{{\sc \emph{Node2Vec}}\xspace}
\newcommand{\deepwalk}{{\sc \emph{DeepWalk}}\xspace}
\newcommand{\versex}{{\sc {Verse}}\xspace}
\newcommand{\mile}{{\sc {Mile}}\xspace}
\newcommand{\harp}{{\sc {Harp}}\xspace}
\renewcommand{\floatpagefraction}{.99}
\title{Understanding Coarsening for Embedding Large-Scale Graphs}


\author{\IEEEauthorblockN{Taha Atahan Akyildiz, Amro Alabsi Aljundi and Kamer Kaya}
\IEEEauthorblockA{ Faculty of Engineering and Natural Sciences, Sabanc{\i}\xspace   University, Istanbul, Turkey\\
E-mail: \tt{\{aakyildiz, amroa, kaya\}@sabanciuniv.edu}}}
\maketitle
\thispagestyle{plain}
\pagestyle{plain}
\pagenumbering{arabic}

\begin{abstract}
A significant portion of the data today, e.g, social networks, web connections, etc., can be modeled by graphs. A proper analysis of graphs with Machine Learning~(ML) algorithms has the potential to yield far-reaching insights into many areas of research and industry. However, the irregular structure of graph data constitutes an obstacle for running ML tasks on graphs such as link prediction, node classification, and anomaly detection. Graph embedding is a compute-intensive process of representing graphs as a set of vectors in a $d$-dimensional space, which in turn makes it amenable to ML tasks. Many approaches have been proposed in the literature to improve the performance of graph embedding, e.g., using distributed algorithms, accelerators, and pre-processing techniques. Graph coarsening, which can be considered a pre-processing step, is a structural approximation of a given, large graph with a smaller one. As the literature suggests, the cost of embedding significantly decreases when coarsening is employed. In this work, we thoroughly analyze the impact of the coarsening quality on the embedding performance both in terms of speed and accuracy. Our experiments with a state-of-the-art, fast graph embedding tool show that there is an interplay between the coarsening decisions taken and the embedding quality. 
\end{abstract}


\section{Introduction}

The data one needs to cope with today for a major discovery is immense, distributed, and unstructured. Although recent advancements in data processing technologies allow us to gather and analyze large-scale data, without scalable algorithms, the data itself does not lend itself to be exploited. An important portion of this data, such as social networks, web connections, financial transactions, co-purchasing information, protein-protein interactions, etc., can naturally be modeled by {\em graphs}. Most of the graphs modeling real-life are  {\em irregular} and {\em sparse}; for instance, in a social network, the number of connections of a person is very small compared to the number of people in the whole network. Furthermore, there is no common pattern that can define all the friends of a given person.\looseness=-1

In the last decade, machine learning (ML) is proven to be disruptive and ML tools and algorithms transformed the way we look and analyze the data and solve our data-centric problems. State-of-the-art ML algorithms are excelled to work on dense and regular (i.e., tabular, image, etc.) data with high precision. Although graphs have the potential to model unique characteristics and complex relations between entities in a data-set very elegantly and efficiently, this, by nature cannot be readily leveraged by ML tools without an explicit transformation which converts the graph to a structure more suitable to be an input to the corresponding ML algorithm. This problem is called {\em graph embedding}. Formally, embedding focuses on representing each vertex of a graph as a vector in a $d$-dimensional space. It allows a wide collection of ML models to be extended to graphs~\cite{linkprediction, deepwalk, anomaly_detection}. Unfortunately, this process is expensive, especially for large, real-life graphs. Many approaches have been proposed to improve the performance of graph embedding, including the usage of distributed systems~\cite{pbg19} and accelerators like GPUs~\cite{graphvite19}. 

Graph coarsening, which is the process of approximating a graph using a smaller, compact one, is investigated to reduce the runtime of the embedding in the literature~\cite{mile18, GOSH20}. These approaches iteratively compress the given graph into smaller, more manageable graphs in a {\em multi-level} setting. At each level, the vertices come together and form a super-vertex, which is simply a set of vertices in the current graph but corresponds to a vertex in the coarsened graph. During this process, some of the edges in the current graph, whose endpoints belong to the same super-vertex set {\em disappear}. After the coarsening levels are completed, the embedding process is iteratively executed on the small, current-level graph and then, the embedding vectors are carried to the previous level. 

As shown in the literature, the effectiveness of coarsening, especially on reducing the embedding runtime is indispensable for large-scale graphs. However, regarding the precision of ML-algorithms,  the properties of a good coarsening strategy and the relative merit of choosing one strategy over another have not been sufficiently examined in the literature. For instance, putting {\em similar} vertices together to create super-vertices is usually a desired coarsening feature, and often the main goal, for the use-cases in graph/hypergraph processing literature such as partitioning~\cite{chaco, scotch, umpa, Catalyurek12, Karypis98}, and community detection~\cite{Blondel_2008, grappolo2}. As we will discuss, multi-level embedding also shares the {\em similarity} desire due to the way it expands the embedding vectors in between the levels.  However, unlike the other use cases, the question "{\em how much similarity is required}" also arises since the disappearing edges are also the connectivity information hidden from the embedding at earlier levels, which is a learning process itself. Hence, the missing connectivity information can only be learned in the later stages of the overall embedding process once the endpoints are uncoarsened. We conjecture that this unique property of graph embedding makes the use of coarsening more interesting and its benefits and exploitation harder to understand. Hence, being unyielding on the similarity and the desire for high-quality coarsening may lead to an inferior learning process for embedding. In this paper, we explore different graph coarsening strategies in terms of their influence on the {\em quality} of graph embedding. The contributions of this paper can be summarized as follows:
\begin{itemize}
    \item To the best of our knowledge, this is the first study solely focusing on the impact of coarsening quality on that of the embedding. An extensive set of experiments using four different coarsening strategies demonstrate that being smart, yet relaxed in terms of similarity has a positive impact on embedding quality.
    \item Recently, we proposed \malgo, a fast, publicly available graph embedding tool leveraging coarsening to exploit the computational power of memory-restricted accelerators such as GPUs. In this work, we analyze the impact of heuristic decisions taken during \malgo's coarsening. We also investigate the impact of coarsening on the embedding quality and show that coarsening not only significantly reduces the runtime but also improves the precision of the ML models for link prediction. 
    \item Not only the coarsening strategy but also how the training budget is distributed among the coarsening levels is important. Our smart work distribution across coarsening levels yield accurate and fast embeddings. On the graph {\tt com-lj}, \malgo can acheive an AUCROC score of $98.7\%$ in the task of link prediction in a single minute where the state-of-the-art~\cite{graphvite19} can reach a similar accuracy in more than 10 minutes. Similarly, the state-of-the-art reports 20 hours of embedding time on multiple GPUs for {\tt com-friendster} which has 60 million vertices and 1.8 billion edges.  On a single GPU and the same graph, our coarsening-based tool reaches $96.6\%$ link prediction AUCROC score in 45 mins.\looseness=-1
\end{itemize}

The rest of the paper is organized as follows: Section~\ref{sec:background} introduces graph embedding and coarsening as well as the notation used in the paper. Section~\ref{sec:methods} describes the coarsening strategies used in the experimentation set. 
Section~\ref{sec:rel} summarizes the existing studies using coarsening for embedding. 
Section~\ref{sec:experiments} presents the experimental results and provides a detailed analysis and inferences about the efficiency and effectiveness of different coarsening approaches. Finally, Section~\ref{sec:conclusion} concludes the paper. 

\section{Background and Notation}\label{sec:background}
A graph $G = (V,E)$ models real-life entities with the vertex set $V$ and their relationships with the edge set $E$. If two vertices $u, v \in V$ have some relationship we have $\{u, v\} \in E$. In this work, we assume that the graph is undirected and the edges are not oriented. Hence, the relationships are bi-directional such as friendship information in a social network or product-buyer information in a purchasing network.   

\subsection{Graph coarsening}

\begin{figure}
    \centering
    \includegraphics[width=0.9\linewidth]{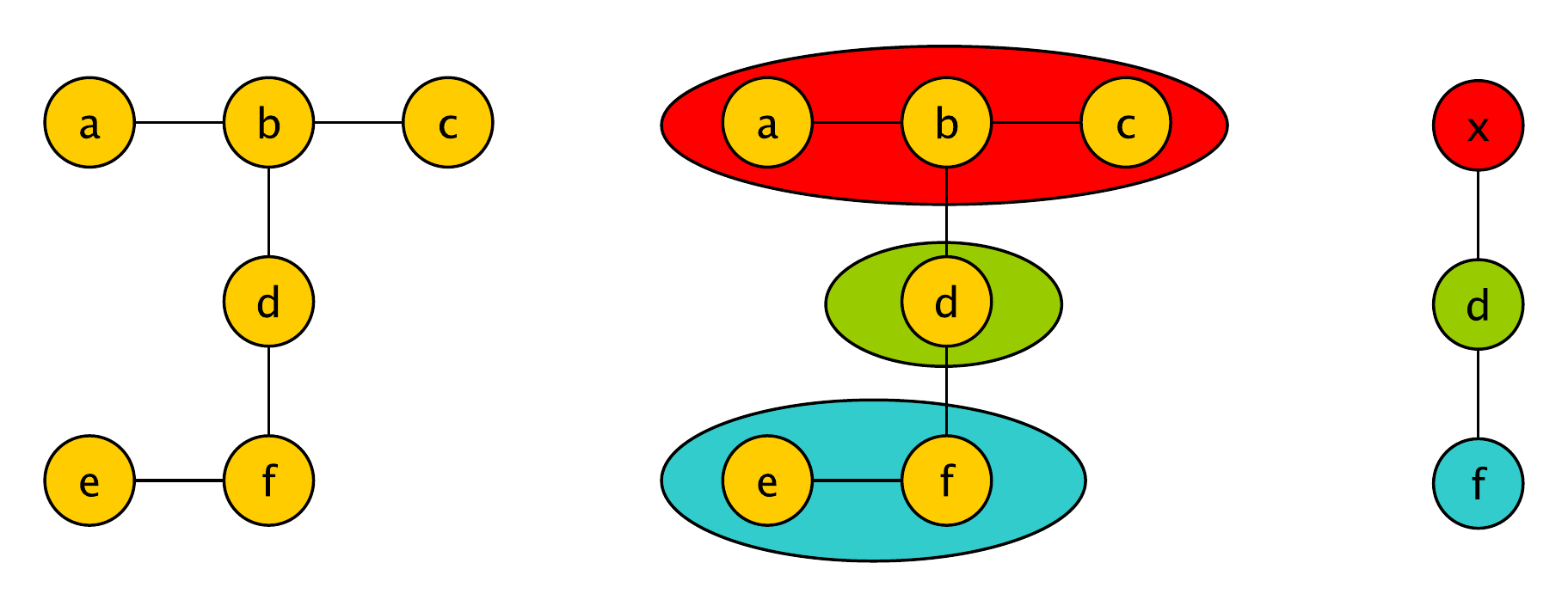}
    \caption{\small{An example of graph coarsening. Groups of one or more vertices in the original graph (left) are grouped based on some predefined coarsening criteria (middle). The coarsened graph (right) is reconstructed by creating super vertices from the grouped vertices and transferring the edge information from the original graph.}}
    \label{fig:coarsening_example}
\end{figure}

Given $G = (V, E)$, graph coarsening is the process of structurally approximating $G$ with a new graph $G' = (V', E')$ such that $G'$ has fewer vertices and edges. This is done through means of collapsing (disjoint) sets of vertices in $G$ into super-vertices which will form the vertex set of $G'$. Figure~\ref{fig:coarsening_example} presents a coarsening step on a toy graph with six vertices and five edges where the coarsened graph has 3 vertices and 2 edges. 

In a multi-level setting, the initial graph $G_0 = G$ is coarsened in multiple levels and a set ${\cal G} = \{G_0, G_1, \cdots, G_{D-1}\}$ of graphs is generated where $G_{D-1}$ is the coarsest, i.e., the smallest graph. 
In this work, we evaluate the \emph{efficiency} of a coarsening level based on the rate of shrinking defined as $${(|V_{i - 1}| - |V_{i}|)}/{|V_{i - 1}|}.$$ We followed a vertex-centric measurement since the size of the embedding matrix and the number of samples required for an iteration change with respect to the number of vertices. We also consider the {\em effectiveness} of the overall coarsening strategy which compares the embedding quality of a strategy to that of another for the same graph embedded with the same parameters.\looseness=-1

\subsection{Graph embedding}
Given a graph $G = (V, E)$, graph embedding  produces an embedding matrix $M \in \mathbb{R}^{|V|\times d}$ where every vertex $v \in V$ is mapped to a $d$-dimensional (row) vector $\mathbf{M}[v]$, which encapsulates $v$'s connectivity information and can be used to carry out many ML tasks. Various embedding methods have been proposed in the last decade~\cite{deepwalk, LINE15, node2vec, verse18, GOSH20}. 

In this paper, we utilize the multi-level, GPU-based embedding tool~\malgo~\cite{GOSH20} as the underlying algorithm to assess the interplay between coarsening strategies and embedding quality. \malgo first iteratively coarsens a graph and obtains $\mathcal{G}$. Then it embeds the coarsest graph $G_{D-1}$, projects its embeddings into the previous level, and carries on the embedding and projection steps until an embedding $\mathbf{M}_{0}$ for the original graph $G_{0}$ is obtained. Projecting the embedding matrix of the graph $G_{i+1}$ to the the graph $G_{i}$ is done by setting the embedding values of every vertex $v \in V_{i}$ to the embedding of its super vertex in $G_{i+1}$. More formally, given that $u, v \in V_{i}$ have been coarsened into the super-vertex $x \in V_{i+1}$, then $\mathbf{M}_{i}[u] = \mathbf{M}_{i}[v] = \mathbf{M}_{i+1}[x]$ where $\mathbf{M}_{i}$ is the embedding matrix of graph $G_{i}$.

Table~\ref{tab:notation} introduces the notation used in the paper.
\begin{table}
\centering
  \caption{\small{Notation used throughout the paper}}
  \label{tab:notation}
  \scalebox{0.9}{
  \begin{tabular}{ll}
    \toprule
    \textbf{Symbol} & \textbf{Definition}\\
    \midrule
    $G_0 = (V_0, E_0)$ & The original graph to be embedded.\\
    $G_{i} = (V_{i}, E_{i})$ & Represents a graph, which is coarsened $i$ times.\\
    $\Gamma^+(u)$ & The set of outgoing neighbors of vertex $u$. \\
    $\Gamma^-(u)$ & The set of incoming neighbors of vertex $u$. \\
    $\Gamma(u)$ & Neighborhood of $u$, i.e., $\Gamma_{G_{i}}^+(u) \bigcup \Gamma_{G_{i}}^-(u)$.\\\hline
    $d$ & $\#$ features per vertex, i.e., dimension of the emb.\\
    $e$ & Total number of epochs that will be performed\\\hline
    $D$ & Total amount of coarsening levels. \\
    ${\mathcal G}$ & The set of coarsened graphs. \\
    $p$ & Smoothing ratio for epoch distribution.\\
    $e_i$ & $\#$ epochs for coarsening level $i$.\\
    ${\mathbf M}_{i}$ & Embedding matrix obtained for $G_{i}$.\\
\end{tabular}}
\end{table}

\section{Understanding Coarsening for Graph Embedding}\label{sec:methods}

As mentioned above, there is an interesting interplay between the coarsening decisions and the performance and quality of the embedding. To understand this we will first dissect the coarsening algorithm of \malgo{}. Then we will create a spectrum of coarsening strategies based on the {\em level of respect} to the vertex similarity. These strategies will be extensively and empirically evaluated later in Section~\ref{sec:experiments}. Last, we will discuss an important step of using coarsening in a multi-level setting, which is distributing the run-time budget to the levels. 

\subsection{Analyzing \malgo coarsening}\label{sec:mec}


To maximize the coarsening {\em efficiency} and {\em effectiveness}, \malgo adapts a coarsening algorithm that favors first-, and second-order proximities~\cite{LINE15} where the former represents the connection between the imminent neighborhood, and the latter represents the similarity between vertices' neighborhoods. The algorithm employs an agglomerative coarsening approach which groups vertices under a super-vertex similar to the approach introduced in~\cite{HARP}. Given a graph $G_{i} = (V_{i}, E_{i})$, each vertex $v \in V_{i}$ is processed one after another. If $v$ is not already mapped under a super-vertex it is marked and mapped under $v_{sup} \in V_{i+1}$. Moreover, $\forall u \in V_{i}$ where $(v, u) \in E_{i}$, if $u$ is not marked, it is also mapped under the cluster $v_{sup}$. Lastly, $G_{i+1}$ is constructed by updating the edges with respect to the newly mapped clusters. With this algorithm, the first-, and second-order proximities are preserved by collapsing vertices around a single vertex.

Performing the coarsening with an arbitrary ordering of vertices constitutes a problem. In the greedy agglomerative coarsening, when a vertex is mapped and added to a cluster, its edges are locked and not processed further. With an arbitrary ordering, vertices with small neighborhoods may lock vertices with relatively larger ones, which in turn does not let the algorithm use the locked edges. Thus, we prefer processing vertices in descending order in terms of their degrees. This is called the {\bf {\em ordering}} heuristic. To be more precise, when two vertices $v$, and $u \in V_{i}$ where $(u, v) \in E_{i}$, and $|\Gamma_{G_{i}}(v)| \geq |\Gamma_{G_{i}}(u)|$ are used for coarsening, $u$ is inserted in to the cluster of origin $v$.

The decisions taken during agglomeration are also important; especially with {\em ordering}, the algorithm demands careful tuning. If two \emph{hub} vertices are mapped under the same super-vertex both the efficiency and the effectiveness of coarsening deteriorate substantially. Note that the probability of this mapping is more compared to mapping lower-degree vertices. From the optimization point of view, such a mapping creates a great mass with a huge gravity force. That is, during embedding at this level, the other vectors will be strongly pulled by the vector corresponding to the super-vertex. If these pulled vertices are not similar, which is usually the case, it will be hard and time consuming to reverse the impact of these optimization steps. These giant hub vertices also inhibit the coarsening process, i.e., they limit the coarsening depth $D$ to a few levels with large numbers of vertices in each level. In order to tackle these problems, a {\bf {\em hub$^2$-restriction}} heuristic is incorporated. That is if both $|\Gamma_{G_{i}}(u)|$ and $|\Gamma_{G_{i}}(v)|$ are larger than  $\frac{|E_{i}|}{|V_{i}|}$, $u$ and $v \in V_{i}$ can not be mapped under the same super-vertex in $V_{i+1}$. The impact of both heuristics will be analyzed in Section~\ref{sec:experiments}. 

\subsection{Distributing the epoch budget}\label{sec:epbud}

The embedding methods in the literature work in epochs, where an epoch is simply a pass over all the vertices. Usually, an epoch budget, i.e., the number of such passes, is given. The budget distribution in \malgo, i.e, the number of epochs assigned to each level, follows a hybrid distribution. A fraction of $p < 1$ epochs are distributed uniformly across levels, while the remaining $(1-p)\times e$ epochs are distributed geometrically such that level $i$ is assigned $e_i = e/D + {e}'_{i}$ epochs where $e'_{i}$ is half of $e'_{i+1}$. Smaller values of $p$ mean that more work is done at the coarser levels which will make the embedding faster, and larger values put more work on the finer levels resulting in more fine-tuned embeddings.  For more details on \malgo coarsening, e.g, time complexity, and parallel implementation details we refer the reader to \cite{GOSH20}.

\subsection{The coarsening spectrum}

Clustering vertices, which share an edge or are similar to each other, is indeed useful since these vertices will have the same embedding vector when multi-level embedding moves up, i.e., uncoarsens, throughout the process. However, if graph topology and proximities are the ultimate, sole information to be exploited, coarsening starts to hide most of the edges in the lower levels. That is, there will be less (positive) information to learn in them. When considering the run-time performance, this is a desired thing to have. However, an important benefit of a multi-level setting is the ability to take big leaps to a good solution which also makes the optimization process effectively jump over local minima. We conjecture that if there exist less information to process, one cannot exploit this property well.

To demystify the impact of coarsening on the quality of the embeddings as much as possible, four coarsening strategies are integrated into \malgo including the default one. All of these strategies merge the vertices in the current level into a super-vertex which forms the center of the corresponding cluster (and appears as a single vertex in the next level). That is in all, the vertices are gathered around a super-vertex to form a cluster. Similarly, all the coarsening approaches have the potential to significantly reduce the training time since the graphs obtained in the multi-level setting are smaller than the original one. However, how they approach the problem, and more formally, how much they have respect for the structure, i.e., the proximity information, during coarsening, are completely different. In fact, they are devised to create a coarsening spectrum and better understand the impact of coarsening on the AUCROC scores. At one hand of the spectrum, there is {\em anti} which always unfavors proximity and coarsens only independent sets. The next one, {\em random}, forms random clusters. The proposed strategy used in \malgo, {\em novel} as called in~\cite{GOSH20}, merges only the neighbor vertices and favors both first and second-order proximities. Hence, it can be placed next to {\em random} in the spectrum. The last strategy, {\em grappolo}, strongly respects the graph structure and proximity information. It forms the clusters and refines them to obtain the {\it best} community structure and maximize the modularity to the most. Hence, by nature, compared to the other three, it hides more edges during the coarsening levels and obtains sparser graphs. It can be placed on the other end of the spectrum. Formally these strategies behave as follows:

\begin{itemize}
    \item {\bf\emph{anti}}: If $(u,v) \in E_{i}$, $u$ and $v$ cannot be mapped under the same super vertex.
    \item {\bf\emph{random}}: If a vertex $v \in V_{i}$ is not yet mapped, first, a vertex $u \in V_{i}$ is selected uniformly randomly. If $u$ is also not mapped, with $v$, they form a super-vertex in $V_{i+1}$. Otherwise, $u$ is re-selected. At most, $|\Gamma(v)|$ selections are executed for each vertex $v$.
    \item {\bf\emph{novel}}: This is the default coarsening strategy used in \malgo as explained above. It respects both the first- and second-order proximities, however, it does not do it aggressively. Furthermore, it avoids large masses that can prematurely end coarsening, i.e produce fewer levels at the end of coarsening.
    \item {\bf\emph{grappolo}}:  Although it has not been directly used for embedding, high-quality, state-of-the-art graph clustering and community detection algorithms have been proposed in the literature. A well-known, CPU-parallel community-detection tool \grappolo~\cite{grappolo2} is selected for further investigating the effect of coarsening on the embeddings.
    The tool is built upon the Louvain method \cite{Blondel_2008} which is an efficient, greedy, and iterative solution for generating a hierarchy of communities~(i.e., clusters). The main idea of Louvain is to maximize the {\em modularity}. A clustering has a high modularity if it has dense intra-cluster connections and sparse inter-cluster connections while keeping the size (i.e., number/weight of the edges) of each cluster balanced. In a multilevel setting, at $i$th level, for $G_{i} = (V_{i}, E_{i})$, the set $P_{i} = \left(C^{(i)}_{1}, C^{(i)}_{2}, \ldots, C^{(i)}_{k}\right)$ represents the communities of $G_{i}$ where $1 \leq k \leq |V_{i}|$. The modularity is calculated as 
\begin{equation}
Q_i = \frac{1}{2m} \sum_{j \in V_{i}} e_{j \rightarrow C^{(i)}(j)} -  \sum_{C \in P_i} \left(\frac{a_{c}}{2m} \times \frac{a_{c}}{2m}\right)
\end{equation}

  where $e_{j \rightarrow C^{(i)}(j)}$ denotes the sum of the edge weights in $E_{j \rightarrow C^{(i)}(j)}$ which is the set of edges that connects vertex $j \in V_i$ to the vertices in its community $C^{(i)}(j)$. The value $a_{C}$ denotes the sum of the edge weights of all the vertices in community $C$~\cite{grappolo2, Newman_2004} and $m$ is the total edge weight in the graph.

\end{itemize}

When combined, the four strategies mentioned above form a nice spectrum; while moving from first to last, the strategies begin to respect more to the topology, become more unyielding and robust, and more restrictive. We use a modularity-based algorithm at one end of this spectrum since it is a common metric to define communities. Also, we use \grappolo since it is a parallel, state-of-the-art tool and able to produce communities with better modularity output compared to the sequential implementation of the Louvain method. Moreover, \grappolo is reported to be $16\times$ faster than the original algorithm with $32$ threads.\looseness=-1 

As explained above, in the multi-level setting, an uncoarsening operation (from level $i$ to $i-1$) initializes the embedding matrix for the current level $(i-1)$ by using the embedding obtained for the coarsened graph from level $i$. For {\em anti}, this yields disjoint sets with totally disconnected vertices whose $d$-dimensional embedding vectors are the same. Hence, most of the epochs for level $(i-1)$ are expected to be spent to pull/push these vertices close to their {\em correct} places. For {\em random}, the picture is almost the same with a few neighbor vertices sharing the same $d$-dimensional vectors. For {\em novel}, each super-vertex in the $i$th level corresponds to a connected component with diameter two in level $(i-1)$ graph. Hence, after the uncoarsening, two vertices can have the same initial embedding vector if and only if their distance on the level $(i-1)$ graph is at most two\footnote{We ignore the cases where two vectors after the embedding at level $i$ are coincidentally the same}. However, if a coarsening decision is not completely correct, which still can be the case, the vertex can slightly be pulled away from its neighbor vertices to a better place since its connections are loose. However, for {\em grappolo}, the force is strong inside the vertex set. Therefore, although the coarsening decisions of {\em grappolo} are probably in concordance with the similarity metric used, they create local minima which are hard to escape from with small learning rates. Hence, we expect that modularity-based coarsening will obtain a good precision even with a few epochs. However, it will probably not be the best.\looseness=-1

\section{Related Work}\label{sec:rel}

To the best of our knowledge, there exist two studies focusing on using coarsening for graph embedding. The first one, MILE~(Multi-Level Embedding Framework), proposes using coarsening to relax computational complexity and overcome memory limitations of the embedding algorithms in the literature~\cite{mile18}. Similar to \malgo, MILE iteratively coarsen the graph. However, it only runs embedding on the smallest graph. Finally, it refines the embedding for the coarsest graph up to the original graph via a Graph Convolutional Neural Network~(GCNN). We refer the reader to ~\cite{kipf2016semisupervised} for more on GCNs.\looseness=-1

MILE applies a {\em hybrid matching}, which includes SEM (Structural Equivalence Matching) and NHEM (Normalized Heavy Edge Matching)~\cite{NHEM}. With SEM, two vertices are matched if and only if they are incident on the same set of neighborhoods, where the vertices are interchangeable and structurally equivalent. With NHEM, an unmatched vertex $u$ is matched with a neighbor $v$ having the largest $(u,v)$ weight. For NHEM, edges are normalized in a fashion where the ones connecting to high-degree vertices are penalized, which in turn prevents forming huge super vertices. To coarsen a graph, SEM is applied first, and all the structurally equivalent vertices are matched. Then the edges are normalized and NHEM is applied. All the matched vertices are collapsed under the respective super-vertices. This process is applied iteratively until the coarsest graph is obtained.

The second related work, \harp (Hierarchical Representation Learning for Networks), is a  meta-strategy proposed to improve graph embedding~\cite{HARP}. Similar to MILE, it recursively coarsens the input graph to get a set of smaller graphs. As in \malgo, after coarsening, \harp runs embedding on all the levels by projecting the vectors of an embedding on a coarsened graph to the embedding of the finer graph. \harp also applies a hybrid coarsening scheme. The scheme has two key parts; edge collapse, and star-collapse for preserving first-order, and second-order proximity respectively. With edge collapse~\cite{hu-drawing}, the vertices that have an edge in between are collapsed such that no vertex can be collapsed more than once. On the other hand, {\em star collapse} matches low-degree peripheral vertices which are connected to the same hub vertex.\looseness=-1



\section{Experiments}\label{sec:experiments}



For the experiments, we use a single server with 2 sockets, each with 8 Intel E5-2620 v4 CPU cores (16 cores in total) running at 2.10GHz. The server has 198GBs RAM. A single Titan X Pascal GPU with 12GB memory is used for GPU experiments.
The operating system on the server is {\tt Ubuntu 4.4.0-159}. {\tt gcc 7.3.0} is used to compile CPU codes with {\tt -O3}, {\tt OpenMP} is used for CPU parallelization. As for GPU implementations and compilation, we used {\tt nvcc} with {\tt CUDA 10.1} and optimization flag {\tt -O3}. The GPUs connection to the server is {\tt PCIe 3.0 x16}. 

\subsection{Tools used for evaluation} 

The following state-of-the-art tools are selected to compare and evaluate the embedding performance:

\begin{itemize}
  \item \versex \footnote{Code publicly available at https://github.com/xgfs/verse}is a generic, multi-core graph embedding algorithm with a fast CPU implementation~\cite{verse18}. We use Personalized Page Rank (PPR) as the similarity measure and $\alpha = 0.85$ as recommended by the authors. The number of epochs and the learning rate (for embedding) are set to $e \in \{600, 1000, 1400\}$ and $lr = 0.0025$, respectively. Out of the three runs, the best AUCROC score is reported.\looseness=-1

    \item \mile \footnote{Code publicly available at https://github.com/jiongqian/MILE} is an embedding algorithm discussed in Section~\ref{sec:rel}. We use {\sc DeepWalk} as the base embedding method, {\sc MD-GCN}~(as suggested in~\cite{mile18}) as the refinement method, $8$ levels of coarsening, and a learning rate of $lr = 0.001$ during embedding. \mile does not allow for the number of epochs to be configured.
  
    \item \malgo \footnote{Code publicly available at https://github.com/SabanciParallelComputing/GOSH} is a fast, GPU-based embedding algorithm equipped with coarsening. We employ four different versions, \malgo-ultra-fast, \malgo-fast, \malgo-normal, and \malgo-slow are used with the parameters given in Table~\ref{table:configs}. We vary the configurations with respect to the number of epochs, the learning rate, and the smoothing ratio (for distributing the epoch budget). As the number of epochs $e$ and the smoothing ratio $p$ decrease, $lr$ increases to compensate for the reduced amount of work on the original graph with faster learning. The only difference between \malgo-ultra-fast, and \malgo-fast is the number of epochs designated for training. In addition to these versions, we use \malgo with no coarsening which spends all the embedding epochs on the original graph. As Table~\ref{table:configs} shows, the configurations use different number of epochs, i.e., $e_{medium}$ and $e_{large}$, for medium-scale and large-scale graphs, respectively. 
\end{itemize}

\begin{table}
\centering
\caption{\small{\malgo configurations: ultra-fast, fast, normal, and small for medium-scale and large-scale graphs. A version with no coarsening is also used in the experiments.}}
\label{table:configs}
\scalebox{1}{
\begin{tabular}{l|rr|r|r}
\textbf{Configuration} & $p$ & $lr$ & $e_{medium}$ & $e_{large}$\\
\midrule
Ultra-fast & 0.1 & 0.050 & 400 & - \\
Fast   & 0.1 & 0.050 & 600 & 100 \\
Normal & 0.3 & 0.035 & 1000 & 200 \\
Slow   & 0.5 & 0.025 & 1400 & 300 \\\hline
No coarsening & - & 0.045 & 1000 & 200  \\
\end{tabular}
}
\end{table}

 In order to make fair comparisons among different graphs, a single epoch is defined as executing $|E|$ number of updates~(as also defined by~\cite{graphvite19}). We do not use \harp in our experiments, since as shown in~\cite{HARP}, the tool is not designed to make the embedding faster and cannot scale to millions of vertices and edges. The main shortcoming is that it does not use the epoch budget distribution described in Section~\ref{sec:epbud} but uses the same, i.e., the original number of epochs at every level. For instance, the largest graph used in~\cite{HARP} has around 10K vertices and less than 350K edges, and for this graph, the reported execution times are all more than 400 seconds. 


\subsection{Datasets} 

For a thorough empirical analysis and to cover different graph structures, we use a wide variety of graphs that have different origins, vertex counts, and densities. The properties of these graphs are shown in Table~\ref{table:graph_summary}. 

\begin{table}
\centering
\caption{\small{The properties of medium-scale (having less than 10M vertices) and large-scale~(having more than 10M vertices) graphs used in the experiments.}} 
\label{table:graph_summary}
\scalebox{0.90}{
\begin{tabular}{lrrrr}
\textbf{Graph}  & $\mathbf{|V|}$ & $\mathbf{|E|}$ & $\mathbf{|E|/|V|}$ & \textbf{Src} \\
\midrule
{\tt com-dblp}  & 317,080      & 1,049,866      & 3.31  &   \cite{snapnets}         \\ 
{\tt com-amazon} & 334,863      & 925,872      & 2.76    &  \cite{snapnets}         \\ 
{\tt youtube}& 1,138,499    & 4,945,382    & 4.34  &  \cite{yt}          \\ 
{\tt soc-pokec}     & 1,632,803 & 30,622,564   &    18.75   &  \cite{snapnets}      \\ 
{\tt wiki-topcats}   & 1,791,489  &  28,511,807  &     15.92 &  \cite{snapnets}       \\ 
{\tt com-orkut}    & 3,072,441    & 117,185,083   &  38.14 &  \cite{snapnets}           \\ 
{\tt com-lj}        & 3,997,962     &  34,681,189  &  8.67   &  \cite{snapnets}         \\ 
{\tt soc-LiveJournal}  & 4,847,571    &  68,993,773  & 14.23 &  \cite{snapnets} \\
\midrule
{\tt hyperlink2012}  & 39,497,204    & 623,056,313       & 15.77       &  \cite{hl}    \\ 
{\tt soc-sinaweibo}   &  58,655,849   &   261,321,071     &  4.46   &   \cite{nr}        \\ 
{\tt twitter\_{rv}}    & 41,652,230    &  1,468,365,182    & 35.25  &   \cite{nr}       \\ 
{\tt com-friendster} & 65,608,366    &   1,806,067,135   &   27.53   &  \cite{snapnets}        \\ 
\end{tabular}
}
\end{table}

\subsection{Evaluation pipeline}
The embedding quality of \malgo, \versex, and \mile are evaluated with link prediction, which is one of the most common machine learning tasks used in the literature to evaluate graph embedding tools~\cite{pbg19, graphvite19, verse18, node2vec}.
 
For evaluation, we split the input graph $G$ into train and test sub-graphs, $G_{train} = (V_{train}, E_{train})$ and $G_{test} = (V_{test}, E_{test})$, respectively. $G_{train}$ contains 80$\%$ of the edges in $G$, and  the remaining 20\% of edges are in $G_{test}$. All isolated vertices in $G_{train}$ are removed. To make sure $V_{test} \subseteq V_{train}$, all $(u,v) \in G_{test}$ edges, where $u$ or $v$ are $\notin G_{train}$ are also removed. Then, the target algorithm is executed with a newly generated $G_{train}$ to get an embedding. Based on this, a Logistic Regression model is trained using the {\tt SGDClassifier} module from {\tt scikit-learn} with a Logistic Regression solver. Finally, the existence of the edges in $G_{test}$ is predicted by the model, and the \emph{AUCROC} score is reported~\cite{roc}.

During the evaluation, two matrices ${\mathbf R}_{train}$ and ${\mathbf R}_{test}$ are created for the prediction pipeline. Each (row) vector ${\mathbf R}_{train}$ is generated by the element-wise multiplication of two vectors ${\mathbf M}[v]$ and ${\mathbf M}[u] \in {\mathbf M}$, where $u, v \in V$. A row is either a positive sample or a negative sample. ${\mathbf R}_{train}$ contains $|E_{train}|$ amount of positive samples, where a positive sample corresponds to an edge $(u,v) \in E_{train}$. Moreover, the same number of negative samples from $(V_{train} \times V_{train}) \setminus E_{train}$ is generated and added as vectors to ${\mathbf R}_{train}$ to produce a balanced data-set for training the logistic regression model. ${\mathbf R}_{test}$ is created similarly by using $G_{test}$ instead of $G_{train}$ as the source of samples. Unlike ${\mathbf R}_{test}$, for the last column of ${\mathbf R}_{train}$, a label representing a positive or negative sample is concatenated to the end of the vector.

\subsection{Experiments on the coarsening spectrum} \label{exp_coarsening_quality}

For embedding, it is hard to demystify which coarsening strategy is the best and why. We devise this experiment to delve into this question and grasp the nature of the optimization led by the decisions taken during the coarsening in a more detailed way. It will not be fair if the strategies are considered competitors. For instance, {\em grappolo} is a modularity maximization tool and not proposed for graph embedding (as {\em novel} is not proposed to maximize modularity). In short, the three other strategies (\emph{anti, random}, and \emph{grappolo}) are judiciously chosen to form the spectrum with {\em novel} which is the default coarsening strategy of \malgo. 

\begin{table}[htbp]
\caption{\small{The performance of \malgo integrated with different types of coarsening strategies. The training graph with a split ratio of $0.8$ is used for both graphs. \malgo-normal is used for the experiments.}}
\label{table:different_types_coarsening}
\centering
\scalebox{0.9} {
\begin{tabular}{lrrrrr}
\textbf{Graph} & \textbf{Strategy} &\textbf{$T_{t}$ (s)} & $D$ & \textbf{$|V_{D - 1}|$} & \textbf{$|E_{D - 1}|$}\\
\midrule
\multirow{4}{*}{{\tt  youtube}} & anti  & 10.3 & 7 & 172 & 29260 \\
& random & 8.3 & 8 & 127 & 15998 \\
& novel & 8.1 & 8 & 202 & 37848 \\
& grappolo & 287.8 & 4 & 10997 & 20039 \\ \midrule
\multirow{4}{*}{{\tt  twitter\_rv}} & anti & 1400.7 & 10 & 279 & 77562 \\
& random & 1195.8 & 11 & 195 & 37584 \\
& novel & 1031.2 & 12 & 158 & 24780 \\
& grappolo & 4586.1 & 4 & 11720 & 40782 \\
\end{tabular}
}
\end{table}

For a glimpse of how these coarsening strategies behave in terms of the number of vertices and edges of the coarsened graphs, one can look at Table~\ref{table:different_types_coarsening}. The first three strategies produce more levels compared to {\em grappolo}. For instance, on {\tt youtube} and {\tt twitter\_rv}, {\em grappolo} takes 4 levels with 20K and 40K edges, on the coarsest graph, respectively. On the other hand, the other three strategies can reach the same number of edges in around 10 levels. As expected, the behavior of {\em anti} and {\em random} are almost similar. As the table shows, for both {\tt youtube} and {\tt twitter\_rv}, {\em random} goes one level deeper and reaches a smaller amount of vertices in the coarsest level. In concordance, the training time for {\em grappolo} is higher, since the epochs are distributed to less number of levels containing relatively larger graphs. 

\begin{figure}[htbp]
    \centering
    \includegraphics[width=0.5\textwidth]{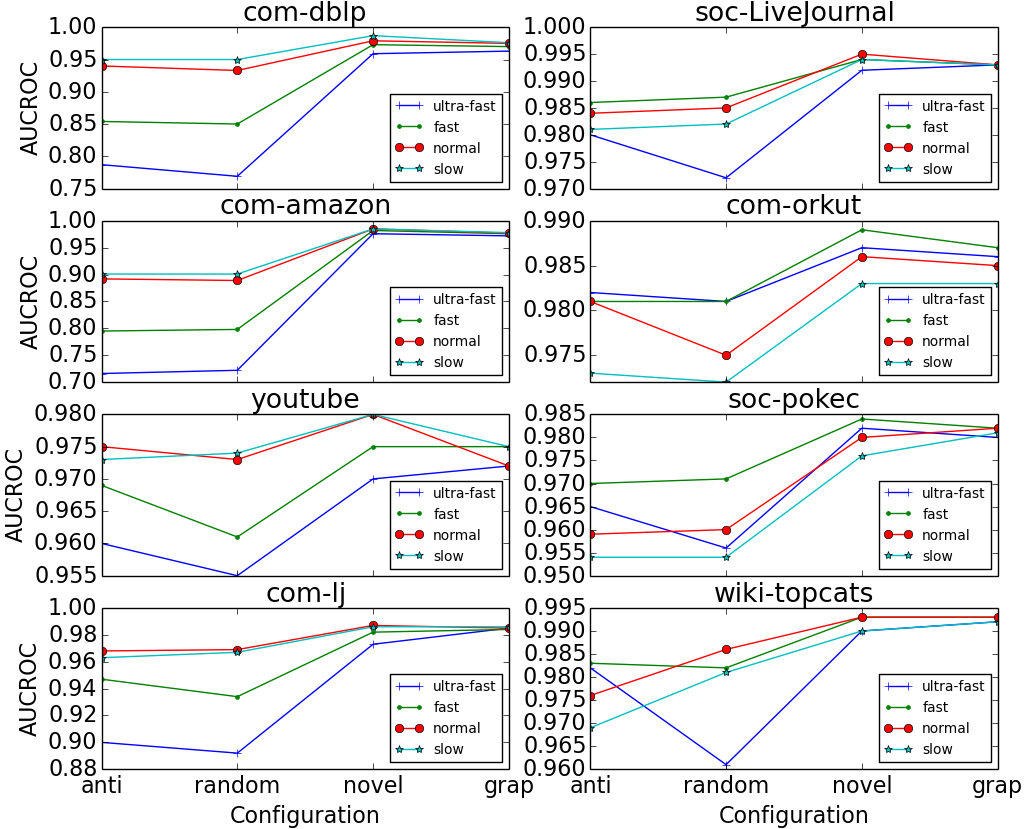}
    \caption{\small{Medium-scale graph results for different coarsening strategies and configurations.}}
    \label{fig:asalgo}
\end{figure}

\begin{figure}[htbp]
    \centering
    \includegraphics[width=0.5\textwidth]{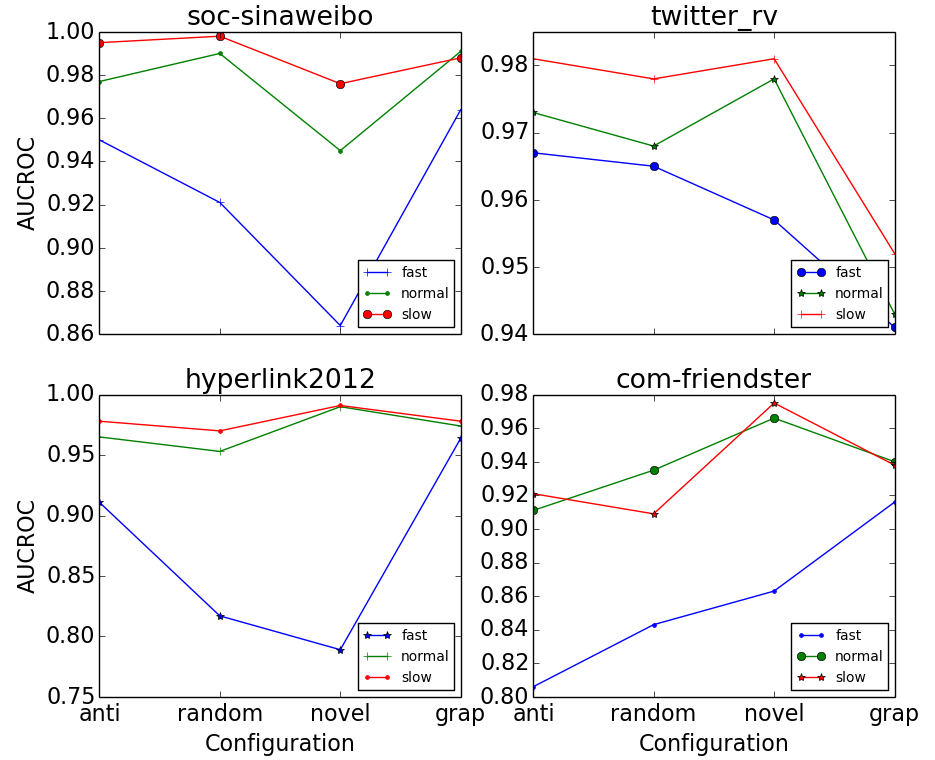}
    \caption{\small{Large-scale graph results for different coarsening strategies and configurations.}}
    \label{fig:abalgo}
\end{figure}

The embedding performance of the aforementioned strategies on different \malgo configurations are given in Figures~\ref{fig:asalgo} and \ref{fig:abalgo} for medium- and large-scale graphs, respectively. In terms of embedding performance, for the configurations ultra-fast and fast, {\em anti} has a slight edge over {\em random}, and for the rest, the performance of both strategies are not distinguishable. For both strategies, the AUCROC scores significantly improve when the number of epochs increases, i.e., when the configurations are changed from ultra-fast to slow. We conjecture that this happens due to fixing the bad decisions taken during coarsening which ignore vertex proximities. Note that with {\em anti}, and also for most of the cases in {\em random}, disconnected vertices will have the same embedding vector at the beginning of each level's embedding. When the number of epochs increases, \malgo has more fuel to fix the negative impact of the underlying coarsening decisions. Although on a smaller scale, such an improvement is also observed for {\em novel} in medium-scale graphs. On these graphs, {\em novel} almost always performs better than {\em anti} and {\em random}. This emphasizes the importance of taking the proximity information into account during coarsening for a better embedding quality.

\begin{figure*}[htbp]
  \centering
  \caption{\small{Performance profile of \malgo using ultra-fast, fast, normal, and slow configurations with different coarsening strategies for the entire data-set. Since the ultra-fast configuration only defined for medium-scale, i.e., 8 out of 12, graphs, the maximum number on the $y$ axis is 8 for this case.}}  \label{fig:kamer}

  \subfloat{\label{fig:kultrafast}\includegraphics[width=45mm]{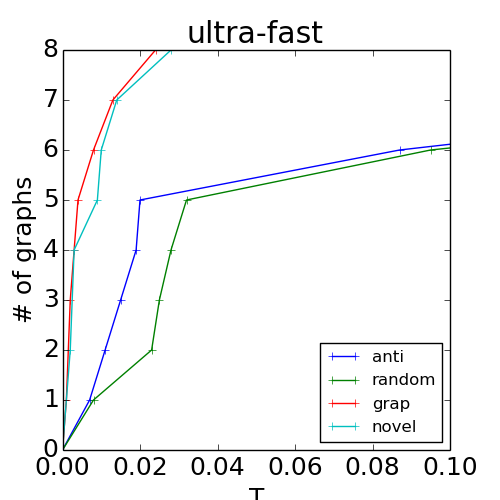}}
  \subfloat{\label{fig:kfast}\includegraphics[width=45mm]{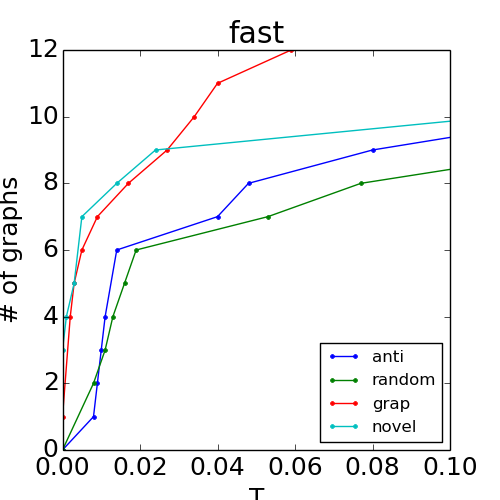}}
  \subfloat{\label{fig:knormal}\includegraphics[width=45mm]{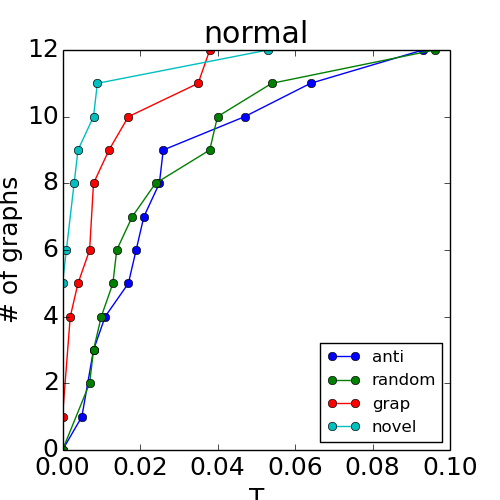}}
  \subfloat{\label{fig:kslow}\includegraphics[width=45mm]{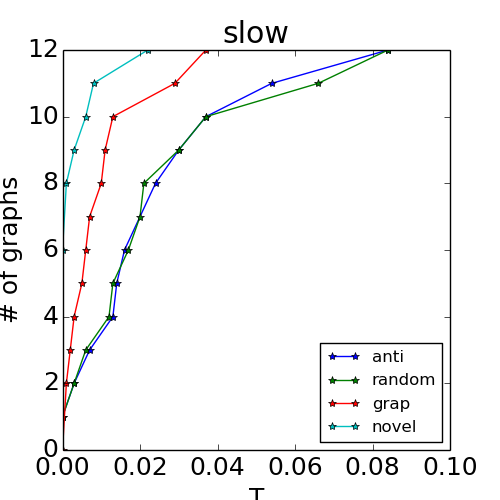}}
\end{figure*}

Especially for large-scale graphs in Figure~\ref{fig:abalgo}, \emph{novel} suffers to generate a quality embedding with a low epoch budget, where \emph{grappolo} scores the best with the exception of {\tt twitter\_rv}. A caveat is that the approaches do not use the same number of levels since \emph{grappolo} produces significantly less coarsening levels. That is the amount of work done by \emph{grappolo} is substantially larger than the rest for fast configurations. 
This being said, it indeed has less AUCROC variation when the configuration and the number of epochs change. When the epoch budget is increased, {\em novel} becomes superior to {\em grappolo} for most of the medium- and large-scale graphs. Since {\em novel} tends to generate smaller graphs with less number of vertices compared to {\em grappolo}, it can perform larger optimization steps, i.e., updates that change the embedding of more vertices (with updates on super-vertices) at once. For instance, on {\tt youtube} with slow configuration, 70 and 220 epochs are reserved  for the original graph by  \emph{novel} and \emph{grappolo}, respectively. For {\em novel}, the final embedding can be considered {\em fine tuning} whereas it is one of the main steps in {\em grappolo}. Furthermore, using more levels also make the embedding process much faster since most of the epochs are spent on smaller graphs. .

\begin{figure}[htbp]
    \centering
    \includegraphics[width=0.5\textwidth]{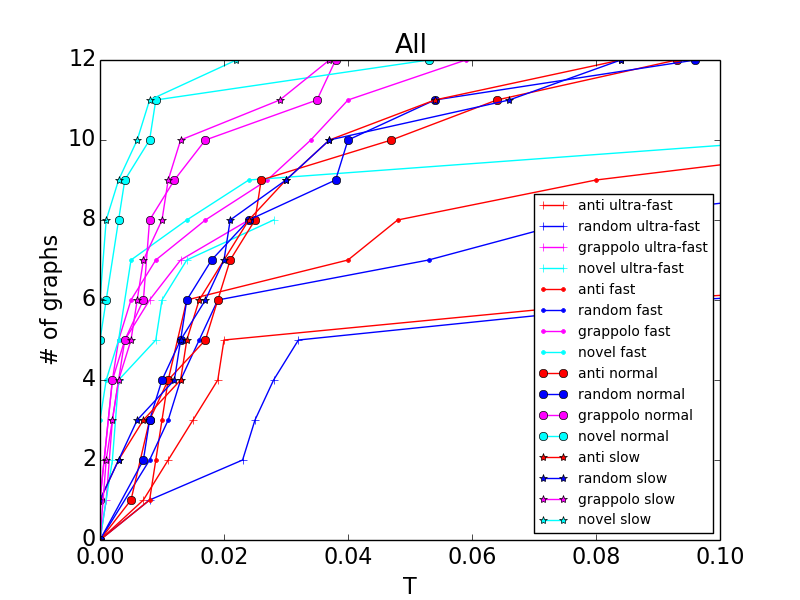}
    \caption{\small{Performance profile of \malgo with different coarsening strategies and embedding configurations for the entire data-set. Colors and markers represent the configuration and the coarsening strategy, respectively.}}
    \label{fig:kall}
\end{figure}

To better profile the relative performance of the coarsening strategies, Figures~\ref{fig:kamer} and~\ref{fig:kall} are provided. In these figures, the algorithms and configurations are compared against the best AUCROC for each graph. The $T$ value ($x$-axis) represents the distance of a strategy to the best \emph{AUCROC} score. A point on the chart, i.e., $T = x$ and $\# of graphs = y$, indicates that for $y$ graphs, the corresponding algorithm scores at most $x$ worse than the best. The previous experiments revealed that {\em grappolo} is the most stable coarsening strategy. That is the variation among its AUCROC scores (with different \malgo configurations) is less. 
In concordance with this, Fig.~\ref{fig:kultrafast}~(left) shows that {\em grappolo} is indeed the best coarsening strategy with the ultra-fast configuration in terms of AUCROC scores. 
We believe that for cheap configurations, the proposed strategy {\em novel} does not have enough number of epochs to fine-tune the embeddings in the last level. However, when \malgo configurations move from ultra-fast to slow~(Fig~\ref{fig:kultrafast} - from left to right), {\em novel} becomes better in terms of precision~(and stays faster than {\em grappolo} in terms of embedding time). Figure~\ref{fig:kall} shows the performance profile of all coarsening-configuration pairs. Similar conclusions follow: {\em anti} and {\em random} need more epochs to obtain a decent performance and when the number of epochs is increased, the change on their AUCROC scores is clear. Although {\em grappolo} is more robust and successful with fewer epochs, {\em novel}, which is a more relaxed strategy, performs better in terms of AUCROC when the number of epochs is increased.

\subsection{Evaluating the impact of heuristics on novel coarsening} \label{sec:coarsening_performance}

As the previous experiments show, coarsening has a huge impact on training time and precision. Hence, it is an indispensable tool to cope with big graphs during embedding. To analyze the performance of the coarsening strategy {\em novel} in detail, we run it {\bf{naive}}ly  without any heuristics, as well as only with {\bf{ordering}}, and in addition, with {\bf hub$^2$-restriction}, both of which are described in Section~\ref{sec:mec}. Table \ref{table:coarsening_optimizations} presents the results of these experiments with these {\em novel} variants. 

\begin{table}[htbp]
\centering
\caption{\small{Performance of \malgo coarsening without any heuristic~({\bf naive}), with {\bf ordering} heuristic, and both with ordering and {\bf hub$^2$-restriction}.}}
\label{table:coarsening_optimizations}
\scalebox{0.90}{
\begin{tabular}{ccrrrrrrr}
 \textbf{} &  & \textbf{AUC} &  & \textbf{} &  &  & \\

 \textbf{} & \textbf{Heur.} & \textbf{ROC} & \textbf{$T_{t}$ (s)} & \textbf{lvl} & \textbf{$T_{c}$ (s)} & $|V_{i}|$ & $\frac{|E_{i}|}{|V_{i}|}$ \\
\midrule
    \multirow{13}{*}{\rotatebox[origin=c]{90}{{\tt youtube}}} 
    & \multirow{2}{*}{{\bf naive}} & \multirow{2}{*}{0.959} & \multirow{2}{*}{106.5} & 1 & 0.13 & 1021590 & 7.8 \\
    & & & & 2 & 0.06 & 636896 & 4.8 \\ \cmidrule{2-8}
    & \multirow{3}{*}{+{\bf ordering}} & \multirow{3}{*}{0.956} & \multirow{3}{*}{79.6} & 1 & 0.17 & 1021590 & 7.8 \\
    & & & & 2 & 0.04 & 480714 & 4.2 \\ 
    & & & & 3 & 0.03 & 378836 & 2.8 \\ \cmidrule{2-8}
    & \multirow{8}{*}{+{\bf hub$^2$-rest.}} & \multirow{8}{*}{0.980} & \multirow{8}{*}{9.3} & 1 & 0.17 & 1021590 & 7.8 \\
    & & & & 2 & 0.04 & 302683 & 12.3 \\ 
    & & & & 3 & 0.02 & 87718 & 32.8 \\
    & & & & 4 & 0.01 & 21356 & 94.3 \\ 
    & & & & 5 & 0.01 & 4436 & 216.3 \\ 
    & & & & 6 & - & 895 & 324.1 \\ 
    & & & & 7 & - & 380 & 298.8 \\
    & & & & 8 & - & 195 & 182.8 \\
\midrule
    \multirow{15}{*}{\rotatebox[origin=c]{90}{{\tt com-friendster}}} 
    & \multirow{2}{*}{{\bf naive}} & \multirow{2}{*}{0.964} & \multirow{2}{*}{43260.0} & 1 & 80.6 & 62603304 & 46.2 \\
    & & & & 2 & 15.99 & 28689906 & 64.2 \\ \cmidrule{2-8}
    & \multirow{2}{*}{+{\bf ordering}} & \multirow{2}{*}{0.921} & \multirow{2}{*}{44912.4} & 1 & 61.99 & 62603304 & 46.2 \\
    & & & & 2 & 9.5 & 26666122 & 44.7 \\ \cmidrule{2-8}
    & \multirow{11}{*}{+{\bf hub$^2$-rest.}} & \multirow{11}{*}{0.972} & \multirow{11}{*}{2316.5} & 1 & 133.09 & 62603304 & 46.2 \\
    & & & & 2 & 98.9 & 20410136 & 131.7 \\ 
    & & & & 3 & 36.69 & 6390812 & 370.2 \\
    & & & & 4 & 7.36 & 1298192 & 842.5 \\ 
    & & & & 5 & 2.02 & 194157 & 1685.4 \\ 
    & & & & 6 & 0.5 & 29415 & 3424.2 \\ 
    & & & & 7 & 0.18 & 8370 & 4094.0 \\
    & & & & 8 & 0.06 & 3421 & 3019.5 \\
    & & & & 9 & 0.01 & 1814 & 1795.1 \\ 
    & & & & 10 & 0.01 & 1100 & 1098.6 \\
    & & & & 11 & - & 823 & 822 \\
\end{tabular}
}
\end{table}

As the table shows, the {\bf naive} variant which allows two vertices with a large degree to merge performs poorly regarding coarsening efficiency. With {\bf ordering} heuristic, a slight improvement in coarsening efficiency is noted. However, the embedding quality degrades substantially; it regresses from $0.964$ to $0.921$ and from $0.959$ to $0.956$ for {\tt com-friendster} and {\tt youtube}, respectively. As explained in Section~\ref{sec:mec}, we believe this is caused by an already existing problem which is amplified by sorting. When two hub-vertices $u,v \in V_{i}$ with large neighborhoods are merged, the coarsening efficiency decreases immensely. However, when the {\bf hub$^2$-restriction} heuristic is added on top of {\bf ordering}, this problem is mitigated. With both heuristics, the largest graph in the dataset {\tt com-friendster} is reduced from 63 million vertices to only 823 vertices in 10 iterations. Furthermore a substantial improvement is also demonstrated regarding coarsening quality; $1\%$, and $2\%$ increase is observed in Table~\ref{table:coarsening_optimizations} for the graphs {\tt com-friendster} and {\tt youtube}, respectively. This shows how these two heuristics complement each other and further highlights the importance of effective and efficient coarsening.

\begin{table}[htbp]
\centering
\caption{\small{The embedding performance with various coarsening depths $D \in \{3, 5, 7\}$.}}
\label{table:coarsening_levels}
\scalebox{0.88} {
\begin{tabular}{l|rr|rr|rr}
\multirow{3}{*}{{\bf Graph}}& \multicolumn{2}{c|}{{$D = 3$}} & \multicolumn{2}{c|}{{$D = 5$}} & \multicolumn{2}{c}{{$D = 7$}} \\
&{\bf T} & {\bf\scriptsize{AUC}}& {\bf T} & {\bf \scriptsize{AUC}}& {\bf T} & {\bf \scriptsize{AUC}}\\
&{\bf (sec)} & {\bf\scriptsize{(\%)}}& {\bf (sec)} & {\bf \scriptsize{(\%)}}& {\bf (sec)} & {\bf \scriptsize{(\%)}}\\
\midrule
{\tt com-dblp} & 11.1 & 97.7 & 5.3 & 97.7 & 3.1 & 97.9 \\
{\tt com-amazon} & 10.5 & 98.0 & 5.1 & 98.4 & 3.0 & 98.5 \\
{\tt youtube} & 45.1 & 97.2 & 20.0 & 97.5 & 11.3 & 98.0 \\
{\tt soc-pokec} & 326.6 & 96.3 & 136.5 & 96.9 & 75.1 & 97.5 \\ 
{\tt com-lj} & 365.3 & 97.3 & 157.5 & 97.8  & 86.6 & 98.5 \\
{\tt com-orkut} & 1117.9 & 97.5  & 474.4 & 97.9  & 263.2 & 98.3 \\
{\tt wiki-topcats} & 226.5 & 98.0  & 95.7 & 98.5  & 54.9 & 99.2 \\

{\tt soc-LiveJour.} & 713.2 & 98.9  & 307.5 & 99.1  & 166.8 & 99.3 \\
\end{tabular}
}
\end{table}

\subsection{Experiments on coarsening depth}\label{sec:depth}
Table~\ref{table:coarsening_levels} shows the impact of using different coarsening levels $D$ on the training time and embedding precision. Due to the epoch distribution strategy of \malgo, when the coarsening depth increases the total amount of work that is reserved for the finer levels decreases. With higher $D$ values, one may expect a decrease both on the training time and the quality of the output, since the amount of epochs transferred from the finer levels to coarser levels will take less amount of time, and the updates on the coarser levels will be less instrumental. While the former is true, as also verified by~\cite{HARP}, the latter is the other way around. As explained in Section~\ref{sec:methods}, when the graph is coarser, the updates on the vertices become more impactful on the finest level. That being said, a good coarsening strategy should not create thousands of levels. In this case, most of the epoch budget will be used for coarsened but similar graphs. Although we do not know the optimal number for each graph, based on our experience and the experiments we present, using around 10 levels creates a good balance between runtime and embedding performance for graphs having 50M vertices and more than 1B edges. 

\subsection{Impact of the coarsening overhead} \label{mileVSgosh}


Although embedding is a time-consuming process, the coarsening overhead can also be important, especially when the embedding tool is fast. For instance, \versex, a CPU-parallel graph embedding tool takes more than 40K seconds on {\tt com-orkut}. It is reported in~\cite{GOSH20} that on the same graph, \mile coarsening and  {\em novel} take 1308.3 and 6.6 seconds, respectively. With \versex, \mile's coarsening overhead can be overlooked. However, \malgo, which is an efficient implementation of \versex on the GPU, only takes 2752 seconds on the same graph without coarsening. Hence, for a fast embedder, the coarsening overhead is important. This is why the attempts such as parallelization to reduce the coarsening overhead without sacrificing from efficiency and effectiveness are valuable to deal with big graphs. 

\begin{table}[htbp]
\centering
\caption{\small{Execution times, number of levels, and the size of the last-level graphs for sequential and parallel coarsening with $\tau = 2, 4, 8, 16$ threads for the large-scale graphs. For {\tt hyperlink2012}, the stopping threshold is set to 0.83 for $\tau = 4, 8$, and $16$.}}
\label{table:seq_parallel_coarsening}
\scalebox{0.90} {
\begin{tabular}{lrrrrrr}
\textbf{Graph}  & $\tau$ &\textbf{Time (s)} & {\bf Speedup} & $D$ & \textbf{$|V_{D - 1}|$} & \textbf{$|E_{D - 1}|$}\\
\midrule
& 1  & 287.8  & -  & 13 & 234 & 53792 \\
& 2 & 224.9 & 1.3$\times$  &   13  & 125 & 15442 \\
{\tt  hyperlink}& 4 & 128.3 & 2.2$\times$ & 13 & 100 & 9884 \\
 {\tt 2012}& 8 & 77.7 & 3.7$\times$ & 13 & 117 & 13560 \\
& 16 & 46.9 & 6.1$\times$ & 13 & 117 & 13572 \\ \midrule
 & 1   & 117.6 & -    & 10 & 230 & 52632  \\
& 2 & 80.4 & 1.3$\times$  &   10  & 187 & 34778 \\
{\tt soc-}& 4 & 46.1 & 2.6$\times$ & 10 & 198 & 39000 \\
{\tt sinaweibo}& 8 & 28.1 & 4.2$\times$ & 10 & 209 & 43472 \\
& 16 & 21.8 & 5.41$\times$ & 9 & 358 & 127716 \\ \midrule
\multirow{5}{*}{{\tt twitter\_rv}} & 1      & 534.9 & -    & 15     & 132 & 17000 \\
& 2 & 400.2 & 1.3$\times$  &   13  & 159 & 24580 \\
& 4 & 226.1 & 2.4$\times$ & 13 & 117 & 13454 \\
& 8 & 131.2 & 4.1$\times$ & 13 & 113 & 12630 \\
& 16 & 77.1 & 6.9$\times$ & 13 & 127 & 16002 \\ \midrule
 & 1  & 2154.5  & -   & 11     & 901 & 810874 \\
& 2 & 1549.4 & 1.4$\times$  &   11  & 768 & 589056 \\
{\tt com-}& 4 & 782.5 & 2.8$\times$ & 11 & 765 & 584460 \\
{\tt friendster}& 8 & 424.9 & 5.1$\times$ & 11 & 788 & 620156 \\
& 16 & 284.5 & 7.6$\times$ & 11 & 795 & 631230 \\
\end{tabular}
}
\end{table}

Table~\ref{table:seq_parallel_coarsening} presents the results of the experiments on sequential and parallel {\em novel} coarsening  with $\tau = 2,4,8,16$ threads. The details of the parallel implementation are given in ~\cite{GOSH20}. As the results show, parallel coarsening has similar $D$ values, and the last-level graphs are of similar sizes. 
Hence, with a similar coarsening quality, the parallel algorithm is $5$--$8\times$ faster compared to the sequential counterpart. As described in ~\cite{GOSH20}, the time complexity of coarsening is ${\mathcal O}$($|V| + |E|$) and in practice, $|E|$ dominates the workload. Although there are other parameters, the variation in the speedups is in concordance with the variation in the number of edges. For instance, {\tt soc-sinaweibo} only has $200$M edges and yields the smallest speedup value of $5.41\times$. On the other hand, the largest speedup $7.57\times$ is obtained for {\tt com-friendster}, which is the largest in our data-set with $1.8$B edges. 

\begin{figure}[htpb]
    \centering
    \includegraphics[width=1\linewidth]{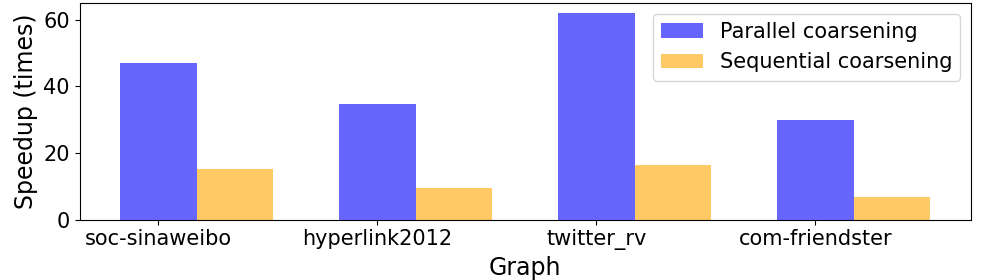}
    \caption{\small{The speedups obtained from running sequentially and parallely coarsened versions of \malgo compared to the non-coarsened version. We used $e = 100$ for all the experiments, i.e., the fast configuration.}}
    \label{fig:speed}
\end{figure}

The impact of using sequential and parallel coarsening on the embedding time can be seen in Figure~\ref{fig:speed}. Indeed, when the fast configuration of \malgo is used, sequential coarsening introduces around $12\times$ speedup on average over the {\em NoCoarse} version of \malgo which spends all the epochs on the original graph. Although this is a significant improvement,  when parallel coarsening is used, the speedup increases to $43\times$. This shows the importance of the attempts to reduce the overhead of coarsening time for graph embedding.

\begin{table}[htbp]
\centering
\caption{\small{Link prediction results on medium-scale graphs. \versex and \malgo uses $\tau = 16$ threads. \mile is a sequential tool. Both \graphvite and \malgo uses the same GPU. The speedup values are computed based on the execution time of \versex.}}
\label{tab:medium_results}
\scalebox{0.90}{
\begin{tabular}{llrrr}
     \toprule
     {\bf Graph} & {\bf Algorithm} & {\bf Time (s)} & {\bf Speedup} & {\bf AUCROC}(\%) \\
     \midrule
     \multirow{7}{5em}{{\tt com-dblp}} & \versex & 248.0 & 1.0$\times$ &  97.8 \\
     & \mile &  136.7 & 1.8$\times$ & 97.7 \\ 
     & \malgo-fast & 0.7  & 335.1$\times$& 98.2 \\
     & \malgo-normal & 2.2  & 110.7$\times$& {\bf 98.5 }\\
     & \malgo-slow & 4.3  & 57.3$\times$& {\bf 98.5} \\
     & \malgo-NoCoarse & 19.5 & 12.7$\times$ & 97.0 \\ 
     \hline
     \multirow{7}{5em}{{\tt youtube}} & \versex & 1365.4 & 1.0$\times$ & \textbf{98.0} \\
     & \mile & 1328.6 & 1.0$\times$ & 94.2 \\
     & \malgo-fast & 2.5 & 557.3$\times$ & 97.5 \\
     & \malgo-normal & 9.3 & 146.3$\times$ & {\bf 98.0} \\
     & \malgo-slow & 19.9 & 68.5$\times$ & {\bf 98.0} \\
     & \malgo-NoCoarse & 108.7 & 8.6$\times$ & 96.7 \\
     \hline
     \multirow{7}{5em}{{\tt com-lj}} & \versex & 12502.7 & 1.0$\times$ & \textbf{98.9}  \\
     & \mile & 3948.6 & 3.2$\times$ & 80.2 \\
     & \malgo-fast & 17.9 & 697.3$\times$ & 98.2 \\
     & \malgo-normal & 62.2 & 201.1$\times$ & {\bf 98.7} \\
     & \malgo-slow & 148.8 & 84.0$\times$ &  98.6 \\
     & \malgo-NoCoarse & 805.5 & 15.5$\times$ & 96.8 \\ 
     \hline
     \multirow{7}{5em}{{\tt com-orkut}} & \versex & 45994.9 &1.0$\times$  &  \textbf{98.7} \\
     & \mile & 11904.3 & 3.9$\times$ & 90.4 \\
     & \malgo-fast & 49.0 & 938.1$\times$ & {\bf 98.9} \\
     & \malgo-normal & 207.1 & 222.1$\times$ & 98.6 \\
     & \malgo-slow & 480.5 & 95.7$\times$ & 98.3 \\
     & \malgo-NoCoarse & 2752.2 & 16.7$\times$ & 96.9 
\end{tabular}
}
\end{table}
\subsection{Performance of \malgo and other tools}
We compare the embedding runtime and precision of \malgo with and without coarsening to those of other tools. Looking at Table~\ref{tab:medium_results}, we can confirm that the embedding quality consistently and significantly improves with coarsening. Furthermore, coarsening helps to catch the AUCROC score of \versex, which is the underlying algorithm for \malgo. We believe that the {\em NoCoarse} version is inferior w.r.t. \versex, since for efficiency, GPU-based embedding is performed in a lock-free fashion which allows some number of race conditions, and hence, incorrect updates on the embedding vectors. Overall, thanks to coarsening, \malgo~(normal) is $110\times$--$222\times$ faster than \versex with a comparable embedding quality. In addition, it is better than \mile both in terms of speed and precision.\looseness=-1

\section{Conclusion and Future Work}\label{sec:conclusion}

Coarsening is shown to be an indispensable technique for graph embedding. It can boost the training performance by several orders of magnitude and enables better utilization of memory-restricted accelerators such as GPUs. Hence, we can assume that it will always be {\em on} for today's high-performance graph embedding tools, as well as the future ones. This is why in this work, we focus on the impact of coarsening strategies over the embedding performance from both computational and optimization points of view. 
 
Although using a strategy that totally follows similarity seems to be robust in terms of precision, we show that its overhead and resistance to fine-grain optimizations can deteriorate the embedding performance in practice. Furthermore, such a strategy saturates the coarsening steps quicker than less restricted strategies which in turn has a negative impact on the embedding time. There exist studies in the literature that focus on improving the runtime of community detection tools. For instance, a recent one improves {\em Grappolo} by an order of magnitude while producing similar or better results~\cite{DBLP:conf/icpp/TithiSAP20}. It is an interesting research avenue to integrate such coarsening strategies to high-performance graph embedding tools in a more suitable way. 


As future work, we would like to extend our analysis to other machine learning tasks, e.g, node classification and anomaly detection. We believe that each task has its own characteristics, and hence it is important to investigate the performance of the coarsening on different ML tasks and explore the possibility of integrating additional metrics to the coarsening. 


\bibliographystyle{IEEEtran}
\bibliography{main}
\end{document}